\documentclass[12pt]{article}
\usepackage{lineno,hyperref}
\usepackage[latin5]{inputenc}
\usepackage{amsmath}
\usepackage{graphicx}
\usepackage{subfigure}
\usepackage{caption}
\usepackage{float}
\usepackage{mathtools}
\usepackage{color}
\usepackage{cite}
\title{Motion of Patterns Modeled by the Gray-Scott Autocatalysis System in One Dimension}
\author{Alper Korkmaz$^{a,}$\thanks{akorkmaz@karatekin.edu.tr}, Ozlem Ersoy$^{b}$, Idiris Dag$^{b}$ \\
$^{a}${\scriptsize Department of Mathematics, Çankırı Karatekin University, 18200, Çankırı, Turkey.} \\
$^{b}${\scriptsize Department of Mathematics \& Computer, Eskisehir Osmangazi University, 26480, Eskisehir, Turkey.}}
\begin{document}
\maketitle
\begin{abstract}

\noindent
Occupation of an interval by self-replicating initial pulses is studied numerically. Two different approximates in different categories are proposed for the numerical solutions of some initial-boundary value problems. The sinc differential quadrature combined with third-fourth order implicit Rosenbrock and exponential B-spline collocation methods are setup to obtain the numerical solutions of the mentioned problems. The numerical simulations containing occupation of single initial pulse, non or slow occupation model and covering the domain with two initial pulses are demonstrated by using both proposed methods.  

\end{abstract}




\section{Introduction}
\noindent
Reaction, diffusion, and reaction-diffusion systems are used to model for many phenomena occuring in physical chemistry and biology\cite{a1,tor,volki1,volki3,nicolis}. According to Nicolis and Prigogine\cite{nicolis}, there exist similarities on evolution of dilute intermediate substances in many cases in both physical chemistry and biology. 
Some at least almost regular patterns frequently form in the concentrations that are reacting chemically and diffusing \cite{nicolis,pear}. Gray-Scott system, which will be described below, is a model for self-replicating structures in two space dimensions and self-replicating pulses in one space dimension\cite{lee,lee2,petrov,reynolds}. Pearson also reported that the spot structures formed by some reaction-diffusion systems clone themselves until the whole domain is covered\cite{pear}. 

\noindent
Consider two irreversible reactions with inert product $P$ and a non-equilibrium constraint represented a feed term for $U$:
\begin{equation*}
\begin{aligned}
U+2V &\xrightarrow{k_1} 3V \\
V &\xrightarrow{k_2} P
\end{aligned}
\end{equation*}
having cubic rate $k_1uv^2$ and linear rate $k_2v$ respectively, and in which both $U$ and $V$ are demolished during the feed process\cite{gray1,hale}. Here, the chemical $U$ reacts with the catalyst $V$ at the rate $k_1$, and the catalyst $V$ decays at a rate $k_2$\cite{hale}. This chemical reaction and diffusion of the self-replicating pattern in one space dimension leads to a nonlinear reaction-diffusion system:
\begin{equation}
\begin{array}{l}
\dfrac{\partial u}{\partial t}=\varepsilon _{1}\dfrac{\partial ^{2}u}{\partial x^{2}}-uv^{2}+f-fu \\ 
\dfrac{\partial v}{\partial t}=\varepsilon _{2}\dfrac{\partial ^{2}v}{\partial x^{2}}+uv^{2}-(f+k)v%
\end{array}
\label{gs}
\end{equation}%
containing dimensionless equations with two chemical concentrations $u(x,t)$ and $v(x,t)$\cite{pear,zeg,reynolds}. In the system, $f$ and $k$ are the dimensionless feed rate and the dimensionless rate constant of the second reaction, respectively, and $\varepsilon _{1}$ and $\varepsilon _{2}$ are diffusion constants. This system is similar to autocatalytic Sel'Kov glycolysis model \cite{selkov} and Gray-Scott model mentioned in\cite{gray2}. Some exact homoclinic and heteroclinic solutions for cubic Gray-Scott autocatalysis system with stability analysis were proposed in \cite{hale}. The linear stability of singular homoclinic and spatially periodic stationary solutions for one dimensional model are analyzed by Doelman et al\cite{doelman}. 
\noindent
In the study, we focus on the solution of initial-boundary value problem in a finite problem interval $[a,b]$ constructed by combining the system\ref{gs} with boundary conditions
\begin{equation}
\begin{aligned}
u(a,t)&=1, \, u(b,t)&=1, \, t\geq 0 \\
v(a,t)&=0, \, v(b,t)&=0, \, t\geq 0 
\end{aligned}
\label{bc}
\end{equation}%
and the initial conditions for both functions given as
\begin{equation}
\begin{aligned}
u(x,0)=g_1(x), \, v(x,0)=g_2(x)
\end{aligned}
\label{ic}
\end{equation}%

\noindent
So far, many types of analytical and numerical solutions have been proposed for many problems arising in the fields covering chemistry, biology, physics and branches of all engineering fields\cite{owo1,kuz1,coo1,akmaz2,sahin1,aksoy1,sahin2,sahin3,akmaz3,geyikli1,akmaz4,esen1}. In many of those methods particularly numerical methods, various techniques were setup by using various basis functions like B-spline functions of different degrees. The structure of those piecewise functions provides easy programmable algorithms, low algebraic calculation costs and reduced complexity. Recently, different from the classical polynomial B-splines, new types of B-spline functions such as extended B-splines have been appeared in some numerical studies\cite{dursun1}.    

\noindent
A particular class of B-spline functions are bell-shaped twicely differentiable exponential cubic B-spline functions and form a basis for the functions defined in the same space\cite{mccartin}. There exist few studies for numerical solutions of linear or nonlinear partial differential equations or systems in the related literature. The collocation method based on exponential B-spline functions was used to solve singular perturbation problem by Sakai and Usmani\cite{sakai}. Radunuvic\cite{radu} developed the collocation method based on cardinal exponential B-splines and solved a singularly perturbed boundary value problem by using the exponential nature and the multiresolution properties of those functions. Another study in the related literature deals with numerical solution of a self-adjoint singularly perturbed Dirichlet boundary value problem by using exponential B-spline functions\cite{rao}. Besides problems for ordinary differential equations, some initial-boundary value problems for linear or nonlinear partial differential equations have also been solved numerically by using combination of exponential B-spline functions and various time integration methods\cite{reza,ozlem,ozlem1}.

\noindent
Different from piecewise B-spline functions, sine cardinal functions have also great importance for many numerical methods  used for solutions of differential equations. Even though those functions have been implemented as basis functions for the solutions of many initial value problems, only one study appears in the related literature which uses them in the differential quadrature method\cite{korkmaz2}.

\noindent
In the present study, we setup an exponential cubic B-spline collocation and sine cardinal functions based differential quadrature methods for solutions of Gray-Scott system. The rest of the paper is organized as follows. Section 2 gives brief descriptions of both sine cardinal functions with their properties and exponential cubic B-spline functions. Then, the design of the proposed methods, discretization of the Gray-Scott system. In Section 3, application of the proposed methods for solutions of initial-boundary value problems for the Gray-Scott system in one dimension. A conclusion is presented in the last section.
\section{Numerical Methods}
\subsection{Sinc Differential Quadrature Method(SDQ) and its Application}

\noindent
The Sinc functions
\begin{equation}
D_{m}(x)=\left\{ 
							\begin{array}{lcc}
										\dfrac{\sin{([\dfrac{x-m\Delta x}{\Delta x}]\pi)}}{[\dfrac{x-m\Delta x}{\Delta x}]\pi} & , & x \neq m\Delta x \\ 
										1 & , & x=m\Delta x \\ 
							\end{array}%
					\right.  \label{sinc}
\end{equation}
where $\Delta x=x_m-x_{m-1}$ is the equal grid size constitute a basis for the functions defined on the real line \cite{stenger,carlson1,secer1,dehghan2}. The functional values of sinc functions at grids are described in \cite{dehghan2} as:
\begin{equation}
D_m(x_j)=\delta_{mj} 
\label{sincnodal}
\end{equation}

\noindent
Let $w$ be defined on ($-\infty ,\infty$). Then, the infinite series
\begin{equation}
C(w)(x)= \sum_{m=-\infty}^{\infty}w(m\Delta x)D_m(x)        
\label{cardinal}
\end{equation} 
is named the cardinal of $w$ if it converges\cite{lund1}. First two principle derivatives of a Sinc function $D_m(x)$ are:
\begin{equation}
D_{m}^{\prime}(x)=\left\{ 
							\begin{array}{lcc}
										\dfrac{\dfrac{\pi}{\Delta x}(x-m\Delta x)\cos{\dfrac{x-m\Delta x}{\Delta x} \pi}-\sin{\dfrac{x-m\Delta x}{\Delta x} \pi} }{\dfrac{\pi}{\Delta x}(x-m\Delta x)^2}& , & x \neq m\Delta x \\ 
										0 & , & x=m\Delta x \\ 
							\end{array}%
					\right.  \label{sincd}
\end{equation}
\begin{equation}
D_{m}^{\prime \prime}(x)=\left\{ 
							\begin{array}{lcc}
							     \dfrac{-\dfrac{\pi}{\Delta x}\sin{\dfrac{x-m\Delta x}{\Delta x} \pi}}{x-m\Delta x}-\dfrac{2\cos{\dfrac{x-m\Delta x}{\Delta x} \pi}}{(x-m\Delta x)^2}+\dfrac{2\sin{\dfrac{x-m\Delta x}{\Delta x} \pi}}{\dfrac{\pi}{\Delta x}(x-m\Delta x)^3}& , & x \neq m\Delta x \\ 
									-\dfrac{{\pi}^2}{3\Delta x^2} & , & x=m\Delta x \\ 
							\end{array}%
					\right.  \label{sincdd}
\end{equation}

\noindent
Differential quadrature derivative approximation model is given as "\textit{the derivative of order $p.$ of a function $w(x)$ at $x_m$ is approximated by finite weighted sum of nodal function values, i.e.,}
\begin{equation}
\left.  \frac{\partial w^{(q)}(x)}{\partial x^{(q)}}\right |_{x=x_{m}}=\sum \limits_{i=1}^{N}c_{mi}^{(q)}w(x_{i}),\quad m=1,2,\ldots ,N,  \label{Funda}
\end{equation}%
\textit{where the partition of the finite problem interval $[a,b]$ is $a=x_1<x_2<\ldots <x_N=b$, $c_{mi}^{(q)}$ are the weighting coefficients of functional values for the $q.$ th order derivative }\cite{bellman1}". The weighting coefficients $c_{mi}^{(q)}$ are determined using any basis function set spanning the problem interval.
\subsubsection{The first order derivative approximation weighting coefficients}

\noindent
Assume that $q=1$ in the fundamental differential quadrature derivative equation\ref{Funda}. The Sinc functions set $\{ D_m(x)\} _{m=1}^{m=N}$ forms a basis for the functions defined on $[x_1=a,b=x_N]$. Then, the weighting coefficients $c_{1i}^{(1)}$ of belonging to the node $x_1$ are determined by substituting each Sinc basis functions $D_m(x)$ into the fundamental differential quadrature equation \ref{Funda}. 
Substitution of $D_{1}(x)$ and using (\ref{sincd}) and (\ref{sincdd}) will lead the algebraic equation
\begin{equation}
\begin{aligned}
D_1^{\prime}(x_1)&=\sum \limits_{i=1}^{N}c_{1i}^{(1)}D_1(x_{i}) \\
     						 &=c_{11}^{(1)}{D_1(x_{1})}+c_{12}^{(1)}{D_1(x_{2})}+\ldots +c_{1N}^{(1)}{D_1(x_{N})} \\
								 &=c_{11}^{(1)}\delta_{11}+c_{12}^{(1)}\delta_{12}+\ldots +c_{1N}^{(1)}\delta_{1N} \\
							0	 &=c_{11}^{(1)} 		 
\end{aligned}
\label{w11}
\end{equation}
to give the weighting coefficient $c_{11}^{(1)}$.

\noindent
The weighting coefficent $c_{12}^{(1)}$ is computed by substitution of $D_{2}(x)$ into Eq.(\ref{Funda}) as
\begin{equation}
\begin{aligned}
D_2^{\prime}(x_1)&=\sum \limits_{i=1}^{N}c_{2i}^{(1)}D_2(x_{i}) \\
     						 &=c_{11}^{(1)}{D_2(x_{1})}+c_{12}^{(1)}{D_2(x_{2})}+\ldots +c_{1N}^{(1)}{D_2(x_{N})} \\
								 &=c_{11}^{(1)}\delta_{21}+c_{12}^{(1)}\delta_{22}+\ldots +c_{1N}^{(1)}\delta_{2N} \\
							 \dfrac{(-1)^{2+1}}{\Delta x (1-2)} &=c_{12}^{(1)} 		 
\end{aligned}
\label{w12}
\end{equation}

\noindent
It is easy to generalize that the weighting coefficients $c_{1i}^{(1)}$ related to the first node $x_1$ are determined by putting each Sinc functions $D_m(x),m=1,2, \ldots , N$ into the differential quadrature approximation equation (\ref{Funda}) as
\begin{equation}
c_{1i}^{(1)}=\dfrac{(-1)^{1-i}}{\Delta x(1-i)},1\neq i  \label{a1i}
\end{equation}%
\begin{equation}
c_{11}^{(1)}=0  \label{a11}
\end{equation}%
When the weighting coefficents $c_{mi}^{(1)}$ belonging to the node $x_m$ is wanted to be determined, a general explicit formulation is used\cite{bellomo1,korkmaz2}:
\begin{equation}
c_{mi}^{(1)}=\dfrac{(-1)^{m-i}}{\Delta x(m-i)},m\neq i  \label{aij}
\end{equation}%
\begin{equation}
c_{mm}^{(1)}=0  \label{aii}
\end{equation}%
\subsubsection{Determination of the second order approximation weights}
When $q$ is assumed as $2$ and $m=1$ in the Eq.(\ref{Funda}), following the same fashion in the determination of the first order derivative approximation weighting coefficients, use of $D_1(x)$ base will lead the equation
\begin{equation}
\begin{aligned}
D_1^{\prime \prime}(x_1)&=\sum \limits_{i=1}^{N}c_{1i}^{(2)}D_1(x_{i}) \\
     						 &=c_{11}^{(2)}{D_1(x_{1})}+c_{12}^{(2)}{D_1(x_{2})}+\ldots +c_{1N}^{(2)}{D_1(x_{N})} \\
								 &=c_{11}^{(2)}\delta_{11}+c_{12}^{(2)}\delta_{12}+\ldots +c_{1N}^{(2)}\delta_{1N} \\
							   \dfrac{-{\pi}^2}{3\Delta x^2}&=c_{11}^{(2)} 		 
\end{aligned}
\label{b12}
\end{equation}
to give $c_{11}^{(2)}$. Substitution of $D_2(x)$ into (\ref{Funda}) will give the equation
\begin{equation}
\begin{aligned}
D_2^{\prime \prime}(x_1)&=\sum \limits_{i=1}^{N}c_{1i}^{(2)}D_2(x_{i}) \\
     						 &=c_{11}^{(2)}{D_2(x_{1})}+c_{12}^{(2)}{D_2(x_{2})}+\ldots +c_{1N}^{(2)}{D_2(x_{N})} \\
								 &=c_{11}^{(2)}\delta_{21}+c_{12}^{(2)}\delta_{22}+\ldots +c_{1N}^{(2)}\delta_{2N} \\
							   2\dfrac{(-1)^{(2+1+1)}}{(\Delta x)^2(1-2)^2}&=c_{12}^{(2)} 		 
\end{aligned}
\label{b22}
\end{equation}
for the weighting coefficient $c_{12}^{(2)}$. Any weighting coefficient $c_{mi}^{(2)}$ focused on the node $x_m$ can be determined by using an explicit formulation \cite{bellomo1,korkmaz2}:
\begin{equation}
c_{mi}^{(2)}=\dfrac{2(-1)^{m-i+1}}{\Delta x^{2}(m-i)^{2}},m\neq i  \label{bij}
\end{equation}%
\begin{equation}
c_{mm}^{(2)}=-\dfrac{\pi ^{2}}{3\Delta x^{2}}  \label{bii}
\end{equation} 
\noindent
\subsubsection{Discretization of the Gray-Scott System}
\noindent
Replacing the space derivative terms by their DQM approximations in Gray-Scott (\ref{gs}) leads to an ordinary differential equation system of the form
\begin{equation}
\begin{aligned}
\left. \frac{\partial u(x,t)}{\partial t} \right |_{x=x_{m}}&=\varepsilon_{1} \sum_{i=1}^{N}{c^{(2)}_{mi}u(x_{i},t)}-u(x_m,t)v^2(x_m,t)+f-fu(x_m,t) \\
\left. \frac{\partial v(x,t)}{\partial t} \right |_{x=x_{m}}&=\varepsilon_{2} \sum_{i=1}^{N}{c^{(2)}_{mi}}v(x_{i},t)+u(x_m,t)v^2(x_m,t)-(f+k)v(x_m,t)\\
\end{aligned} \label{discdqm}
\end{equation}
where $c_{mi}^{(2)}$ are the weighting coefficients of each $u(x_{m},t)$ and $v(x_{m},t)$ for the second order derivative approximations at the grid $x_m$. Before time integration of the system \ref{discdqm}, the implementation of the boundary conditions reduces it to 
\begin{equation}
\begin{aligned}
\left. \frac{\partial u(x,t)}{\partial t} \right |_{x=x_{m}}&=\varepsilon_{1} \sum_{i=2}^{N-1}{c^{(2)}_{mi}u(x_{i},t)}-u(x_m,t)v^2(x_m,t)+f-fu(x_m,t)+c^{(2)}_{m1}+c^{(2)}_{mN} \\
\left. \frac{\partial v(x,t)}{\partial t} \right |_{x=x_{m}}&=\varepsilon_{2} \sum_{i=2}^{N-1}{c^{(2)}_{mi}}v(x_{i},t)+u(x_m,t)v^2(x_m,t)-(f+k)v(x_m,t)\\
\end{aligned} \label{discdqm2}
\end{equation}
The fully space discretized system (\ref{discdqm2}) is integrated with respect to the time variable $t$ by using the third-fourth order Runge-Kutta Rosenbrock method.

\subsection{Exponential Cubic B-spline Collocation Method(ECC)}
\begin{figure}[htbp]
	\centering
		\includegraphics[scale=0.4]{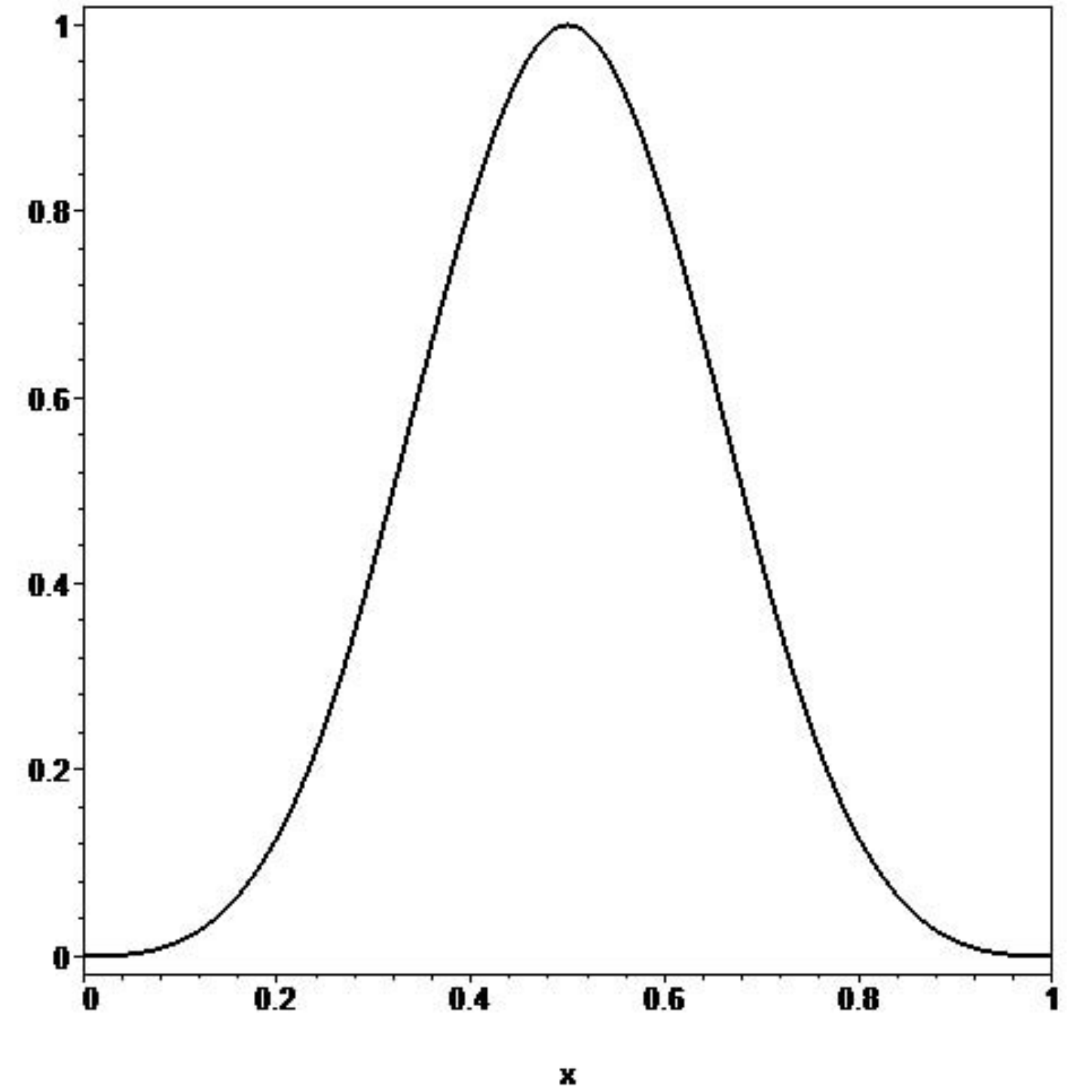}
	\caption{Exponential B-spline for $\lambda=1$}
	\label{fig:Fig1}
\end{figure}

\noindent
Let $\pi $ be a uniform partition of the finite interval $[a,b]$ defined as: 
\begin{equation*}
\pi :a=x_{0}<x_{1}<\ldots <x_{N}=b
\end{equation*}%
with equal mesh size $h=x_{m+1}-x_m\, , m=0,1, \ldots , N-1$.
Then, the exponential cubic B-spline functions are defined as;
\begin{equation}
C_{m}(x)=\left\{ 
\begin{array}{lcc}
b_{2}\left ( (x_{m-2}-x)-\frac{1}{\lambda} \sinh(\lambda(x_{m-2}-x)) \right ) & , & [x_{m-2},x_{m-1}] \\ 
a_{1}+b_{1}(x_{m}-x)+c_{1}\exp(\lambda(x_{m}-x))+d_{1}\exp(-\lambda(x_{m}-x)) & , & [x_{m-1},x_{m}] \\ 
a_{1}+b_{1}(x-x_{m})+c_{1}\exp(\lambda(x-x_{m}))+d_{1}\exp(-\lambda(x-x_{m})) & , & [x_{m},x_{m+1}] \\ 
b_{2}\left ( (x-x_{m+2})-\frac{1}{\lambda} \sinh(\lambda(x-x_{m+2})) \right ) & , & [x_{m+1},x_{m+2}] \\ 
0 & , & otherwise%
\end{array}%
\right.  \label{ecbs}
\end{equation}%
where 
$$
\begin{array}{l}
a_{1}=\dfrac{\lambda h\cosh(\lambda h)}{\lambda h\cosh(\lambda h)-\sinh(\lambda h)}, \\
b_{1}=\dfrac{\lambda }{2} \dfrac{\cosh(\lambda h)(\cosh(\lambda h)-1)+\sinh^2(\lambda h)}{(\lambda h\cosh(\lambda h)-\sinh(\lambda h))(1-\cosh(\lambda h))}, \\
b_{2}=\dfrac{\lambda }{2(\lambda h\cosh(\lambda h)-\sinh(\lambda h))}, \\
c_{1}=\dfrac{1}{4} \dfrac{\exp(-\lambda h)(1-\cosh(\lambda h))+\sinh(\lambda h)(\exp(-\lambda h)-1))}{(\lambda h\cosh(\lambda h)-\sinh(\lambda h))(1-\cosh(\lambda h))}, \\
d_{1}=\dfrac{1}{4} \dfrac{\exp(\lambda h)(\cosh(\lambda h)-1)+\sinh(\lambda h)(\exp(\lambda h)-1))}{(\lambda h\cosh(\lambda h)-\sinh(\lambda h))(1-\cosh(\lambda h))},
\end{array}
$$ 
where $\lambda $ is a real parameter\cite{mccartin}. The exponential B-spline function set $\{C_{m}(x)\}_{m=-1}^{m=N+1}$ constitutes a basis for the functions defined over the interval $[a,b]$. The bell-shape of the exponential cubic B-spline function for $\lambda=1$ is demonstrated in Fig. \ref{fig:Fig1}.

\noindent
The nonzero functional values of each exponential B-spline $C_{m}(x)$ and its two principle derivatives at the grids are tabulated in Table \ref{table:ebs}.
\scriptsize{
\begin{table}[h]
\caption{$C_{m}(x)$ and its principle two derivatives at the grids}
\scriptsize{
\begin{tabular}{lccccc}
\hline \hline
$x$ & $x_{m-2}$ & $x_{m-1}$ & $x_{m}$ & $x_{m+1}$ & $x_{m+2}$ \\ 
\hline \\
$C_{m}$ & $0$ & $\dfrac{\sinh(\lambda h)-\lambda h}{2(\lambda h\cosh(\lambda h)-\sinh(\lambda h))}$ & $1$ & $\dfrac{\sinh(\lambda h)-\lambda h}{2(\lambda h\cosh(\lambda h)-\sinh(\lambda h))}$ & $%
0$ \\ 
$C_{m}^{^{\prime }}$ & $0$ & $\dfrac{\lambda (1-\cosh(\lambda h))}{2(\lambda h\cosh(\lambda h)-\sinh(\lambda h))}$ & $0$ & $\dfrac{%
\lambda (\cosh(\lambda h)-1)}{2(\lambda h\cosh(\lambda h)-\sinh(\lambda h))}$ & $0$ \\ 
$C_{m}^{^{\prime \prime }}$ & $0$ & $\dfrac{\lambda ^{2}\sinh(\lambda h)}{2(\lambda h\cosh(\lambda h)-\sinh(\lambda h))}$ & $-\dfrac{%
\lambda ^{2}\sinh(\lambda h)}{\lambda h\cosh(\lambda h)-\sinh(\lambda h)}$ & $\dfrac{\lambda ^{2}\sinh(\lambda h)}{2(\lambda h\cosh(\lambda h)-\sinh(\lambda h))}$ & $0$ \\ 
\hline \hline
\end{tabular}}
\label{table:ebs}
\end{table}
\normalsize

\noindent
Assume that the solutions $u(x,t)$ and $v(x,t)$ be in the form
\begin{equation}
{u}(x,t)\cong\sum_{i=-1}^{N+1}\delta _{i}C_{i}(x),\text{ \  \  \ }%
{v}(x,t)\cong\sum_{i=-1}^{N+1}\phi _{i}C_{i}(x)  \label{g1}
\end{equation}%
where $\delta _{i}$ are time dependent variables that will be determined from the collocation method. The first two principle derivatives of ${u}(x,t)$ and $%
{v}(x,t)$ can be determined as 
\begin{equation}
\begin{aligned}
&{u}^{\prime }(x,t)\cong\sum_{i=-1}^{N+1}\delta _{i}C_{i}^{\prime }(x) ,\,\,
&{u}^{\prime \prime}(x,t)\cong\sum_{i=-1}^{N+1}\delta _{i}C_{i}^{\prime\prime }(x)\\
&{v}^{\prime}(x,t) \cong\sum_{i=-1}^{N+1}\phi _{i}C_{i}^{\prime}(x),\,\,
&{v}^{\prime \prime}(x,t)\cong\sum_{i=-1}^{N+1}\phi _{i}C_{i}^{\prime\prime }(x) \label{der}
\end{aligned}
\end{equation}

\noindent
Using the equations (\ref{g1}) and (\ref{der}) with functional values of each exponential B-spline given in
Table \ref{table:ebs}, and its first two principle derivatives at the grids can be written in the form
\begin{equation}
\begin{array}{c}
\begin{tabular}{l}
${u}_{i}={u}(x_{i},t)\cong\dfrac{s-\lambda h}{2(\lambda hc-s)}\delta _{i-1}+\delta _{i}+\dfrac{s-\lambda h%
}{2(\lambda hc-s)}\delta _{i+1}$ \\ 
${u}_{i}^{\prime }={u}^{\prime }(x_{i},t)\cong\dfrac{\lambda (1-c)}{2(\lambda hc-s)}\delta _{i-1}+%
\dfrac{\lambda (c-1)}{2(\lambda hc-s)}\delta _{i+1}$ \\ 
${u}_{i}^{\prime \prime }={u}^{\prime \prime }(x_{i},t)\cong\dfrac{\lambda ^{2}s}{2(\lambda hc-s)}%
\delta _{i-1}-\dfrac{\lambda ^{2}s}{\lambda hc-s}\delta _{i}+\dfrac{\lambda ^{2}s}{2(\lambda hc-s)}%
\delta _{i+1}.$%
\end{tabular}
\\ 
\begin{tabular}{l}
${v}_{i}={v}(x_{i},t)\cong\dfrac{s-\lambda h}{2(\lambda hc-s)}\phi _{i-1}+\phi _{i}+\dfrac{s-\lambda h}{%
2(\lambda hc-s)}\phi _{i+1},$ \\ 
${v}_{i}^{\prime }={v}^{\prime }(x_{i},t)\cong\dfrac{\lambda (1-c)}{2(\lambda hc-s)}\phi _{i-1}+%
\dfrac{\lambda (c-1)}{2(\lambda hc-s)}\phi _{i+1}$ \\ 
${v}_{i}^{\prime \prime }={v}^{\prime \prime }(x_{i},t)\cong\dfrac{\lambda ^{2}s}{2(\lambda hc-s)}%
\phi _{i-1}-\dfrac{\lambda ^{2}s}{\lambda hc-s}\phi _{i}+\dfrac{\lambda ^{2}s}{2(\lambda hc-s)}\phi
_{i+1}.$%
\end{tabular}%
\end{array}
\label{g4}
\end{equation}
where $s=\sinh{\lambda h}$ and $c=\cosh{\lambda h}$. After discretizing the Gray-Scott system (\ref{gs}) in time by the Crank-Nicholson method, it reduces to 
\begin{equation}
\begin{aligned}
\dfrac{{u}^{n+1}-{u}^{n}}{\Delta t}&=\varepsilon _{1}\dfrac{%
{u}_{xx}^{n+1}+{u}_{xx}^{n}}{2}-\dfrac{({u}^{2}{v})^{n+1}+({u}^{2}{v})^{n}}{2}+f(1-%
\dfrac{{u}^{n+1}+{u}^{n}}{2}) \\ 
\dfrac{{v}^{n+1}-{v}^{n}}{\Delta t}&=\varepsilon _{2}\dfrac{%
{v}_{xx}^{n+1}+{v}_{xx}^{n}}{2}+\dfrac{({u}^{2}{v})^{n+1}+({u}^{2}{v})^{n}}{2}+(f+k)(%
\dfrac{{v}^{n+1}+{v}^{n}}{2})%
\end{aligned}
\label{g5}
\end{equation}%
where ${u}^{n+1}={u}(x,t^{n+1})$ and ${v}^{n+1}={v}(x,t^{n+1})$ denote the solutions of Gray-Scott system at the $(n+1).$th time level. It should be mentioned that here $t^{n+1}=t^{n}+\Delta t$, and $t^{n}=n\Delta t$. Substitution of approximate solutions into (\ref{g5}) and rearranging the resultant system lead to

\begin{eqnarray}
&&\nu _{m1}\delta _{m-1}^{n+1}+\nu _{m2}\phi _{m-1}^{n+1}+\nu _{m3}\delta
_{m}^{n+1}+\nu _{m4}\phi _{m}^{n+1}+\nu _{m1}\delta _{m+1}^{n+1}+\nu
_{m2}\phi _{m+1}^{n+1}  \label{f1} \\
&=&\nu _{m5}\delta _{m-1}^{n}+\nu _{m6}\delta _{m}^{n}+\nu _{m5}\delta
_{m+1}^{n}  \notag
\end{eqnarray}%
and%
\begin{eqnarray}
&&\nu _{m7}\delta _{m-1}^{n+1}+\nu _{m8}\phi _{m-1}^{n+1}+\nu _{m9}\delta
_{m}^{n+1}+\nu _{m10}\phi _{m}^{n+1}+\nu _{m7}\delta _{m+1}^{n+1}+\nu
_{m8}\phi _{m+1}^{n+1}  \label{f2} \\
&=&\nu _{m11}\phi _{m-1}^{n}+\nu _{m12}\phi _{m}^{n}+\nu _{m11}\phi
_{m+1}^{n}  \notag
\end{eqnarray}%
where%
\begin{equation*}
\begin{array}{lll}
\nu _{m1}=\left( \dfrac{2}{\Delta t}+f+L_{1}^{2}\right) \alpha
_{1}-\varepsilon _{1}\gamma _{1} &  & \nu _{m7}=L_{1}^{2}\alpha _{1} \\ 
\nu _{m2}=2K_{1}L_{1}\alpha _{1} &  & \nu _{m8}=\left( \dfrac{2}{\Delta t}%
+(f+k)+2K_{1}L_{1}\right) \alpha _{1}-\varepsilon _{2}\gamma _{1} \\ 
\nu _{m3}=\left( \dfrac{2}{\Delta t}+f+L_{1}^{2}\right) -\varepsilon
_{1}\gamma _{2} &  & \nu _{m9}=L_{1}^{2} \\ 
\nu _{m4}=2K_{1}L_{1} &  & \nu _{m10}=\left( \dfrac{2}{\Delta t}%
+(f+k)+2K_{1}L_{1}\right) -\varepsilon _{2}\gamma _{2} \\ 
\nu _{m5}=\left( \dfrac{2}{\Delta t}-f+L_{1}^{2}\right) \alpha
_{1}+\varepsilon _{1}\gamma _{1} &  & \nu _{m11}=\left( \dfrac{2}{\Delta t}%
-(f+k)-K_{1}L_{1}\right) \alpha _{1}+\varepsilon _{2}\gamma _{1} \\ 
\nu _{m6}=\left( \dfrac{2}{\Delta t}-f+L_{1}^{2}\right) +\varepsilon
_{1}\gamma _{2} &  & \nu _{m12}=\left( \dfrac{2}{\Delta t}%
-(f+k)-K_{1}L_{1}\right) +\varepsilon _{2}\gamma _{2}%
\end{array}%
\end{equation*}%
\begin{equation*}
\begin{array}{cc}
K_{1}=\alpha _{1}\delta _{m-1}^{n}+\alpha _{2}\delta _{m}^{n}+\alpha
_{3}\delta _{m+1}^{n} & L_{1}=\alpha _{1}\phi _{m-1}^{n}+\alpha _{2}\phi
_{m}^{n}+\alpha _{3}\phi _{m+1}^{n}%
\end{array}%
\end{equation*}%
\begin{equation*}
\alpha _{1}=\dfrac{s-\lambda h}{2(\lambda hc-s)},\text{ }\gamma _{1}=\dfrac{\lambda ^{2}s}{%
2(\lambda hc-s)},\text{ }\gamma _{2}=-\dfrac{\lambda ^{2}s}{phc-s}.
\end{equation*}

\noindent
The system with (\ref{f1}) and (\ref{f2}) can be converted the following
matrices system;%
\begin{equation}
\mathbf{Ax}^{n+1}=\mathbf{Bx}^{n}+2\mathbf{C}  \label{f3}
\end{equation}%
where%
\begin{equation*}
\mathbf{A=}%
\begin{bmatrix}
\nu _{m1} & \nu _{m2} & \nu _{m3} & \nu _{m4} & \nu _{m1} & \nu _{m2} &  & 
&  &  \\ 
\nu _{m5} & 0 & \nu _{m6} & 0 & \nu _{m5} & 0 &  &  &  &  \\ 
&  & \nu _{m1} & \nu _{m2} & \nu _{m3} & \nu _{m4} & \nu _{m1} & \nu _{m2} & 
&  \\ 
&  & \nu _{m5} & 0 & \nu _{m6} & 0 & \nu _{m5} & 0 &  &  \\ 
&  &  & \ddots & \ddots & \ddots & \ddots & \ddots & \ddots &  \\ 
&  &  &  & \nu _{m1} & \nu _{m2} & \nu _{m3} & \nu _{m4} & \nu _{m1} & \nu
_{m2} \\ 
&  &  &  & \nu _{m5} & 0 & \nu _{m6} & 0 & \nu _{m5} & 0%
\end{bmatrix}%
\end{equation*}%
\begin{equation*}
\mathbf{B=}%
\begin{bmatrix}
\nu _{m7} & \nu _{m8} & \nu _{m9} & \nu _{m10} & \nu _{m7} & \nu _{m8} &  & 
&  &  \\ 
0 & \nu _{m11} & 0 & \nu _{m12} & 0 & \nu _{m11} &  &  &  &  \\ 
&  & \nu _{m7} & \nu _{m8} & \nu _{m9} & \nu _{m10} & \nu _{m7} & \nu _{m8}
&  &  \\ 
&  & 0 & \nu _{m11} & 0 & \nu _{m12} & 0 & \nu _{m11} &  &  \\ 
&  &  & \ddots & \ddots & \ddots & \ddots & \ddots & \ddots &  \\ 
&  &  &  & \nu _{m7} & \nu _{m8} & \nu _{m9} & \nu _{m10} & \nu _{m7} & \nu
_{m8} \\ 
&  &  &  & 0 & \nu _{m11} & 0 & \nu _{m12} & 0 & \nu _{m11}%
\end{bmatrix}%
\text{ and }\mathbf{C=}%
\begin{bmatrix}
f \\ 
0 \\ 
\vdots \\ 
f \\ 
0%
\end{bmatrix}%
\end{equation*}

\noindent
The system (\ref{f3}) consists of $2N+2$ linear equations with $2N+6$ unknown
parameters $\mathbf{x}^{n}=(\delta _{-1}^{n},\phi _{-1}^{n},\delta
_{0}^{n},\phi _{0}^{n},\ldots ,\delta _{N+1}^{n},\phi _{N+1}^{n})$. Adapting the boundary conditions to new variables $\delta$ and $\phi$ and eliminating the variables with subscripts $-1$ and $N+1$ by substituting 

\begin{equation}
\begin{array}{l}
\delta _{-1}^{n}=\dfrac{2(\lambda hc-s)}{s-\lambda h}-\dfrac{2(\lambda hc-s)}{s-\lambda h}\delta
_{0}^{n}-\delta _{1}^{n}, \\ 
\phi _{-1}^{n}=-\dfrac{2(\lambda hc-s)}{s-\lambda h}\phi _{0}^{n}-\phi _{1}^{n}, \\ 
\delta _{N+1}^{n}=\dfrac{2(\lambda hc-s)}{s-\lambda h}-\delta _{N-1}^{n}-\dfrac{2(\lambda hc-s)}{%
s-\lambda h}\delta _{N}^{n}, \\ 
\phi _{N+1}^{n}=-\phi _{N-1}^{n}-\dfrac{2(\lambda hc-s)}{s-\lambda h}\phi _{N}^{n},%
\end{array}
\label{LBC}
\end{equation}
into the system (\ref{f3}) generates a solvable system containing equal equations and number of unknowns.
The reduced septa-diagonal system of $(2N+2)$ equations with $(2N+2)$ is solved by well known Thomas algorithm with inner iteration to improve the results in each time step.

\subsection{Adapting the Initial State}
\noindent
In order to be able to start time integration of the iterative system (\ref{f3}), the initial parameters $\delta _{-1}^{0},$ $\phi _{-1}^{0},$ $\delta_{0}^{0},$ $\phi _{0}^{0},\ldots ,\delta _{N+1}^{0},$ $\phi _{N+1}^{0}$ are required to be calculated from the initial condition and first space derivative of the initial conditions at the boundaries as the following%

\begin{equation}
\begin{aligned}
{u}^{^{\prime }}(a,0)&=\dfrac{\lambda(1-c)}{2(\lambda hc-s)}\delta^{0} _{-1}+\dfrac{\lambda (c-1)}{%
2(\lambda hc-s)}\delta^{0} _{1} \\ 
{u}(x_{m},0)&=\dfrac{s-\lambda h}{2(\lambda hc-s)}\delta _{m-1}^{0}+\delta _{m}^{0}+\dfrac{%
s-\lambda h}{2(\lambda hc-s)}\delta _{m+1}^{0},\, m=0,1,..,N-1 \\ 
{u}^{\prime }(b,0)&=\dfrac{\lambda (1-c)}{2(\lambda hc-s)}\delta^{0} _{N-1}+\dfrac{\lambda (c-1)}{%
2(\lambda hc-s)}\delta^{0} _{N+1}%
\end{aligned}
\label{B1}
\end{equation}%
and%
\begin{equation}
\begin{aligned}
{v}^{^{\prime }}(a,0)&=\dfrac{\lambda (1-c)}{2(\lambda hc-s)}\phi^{0} _{-1}+\dfrac{\lambda (c-1)}{%
2(\lambda hc-s)}\phi^{0} _{1} \\ 
{v}(x_{m},0)&=\dfrac{s-\lambda h}{2(\lambda hc-s)}\phi _{m-1}^{0}+\phi _{m}^{0}+\dfrac{s-\lambda h}{%
2(\lambda hc-s)}\phi _{m+1}^{0},\, m=0,1,...,N \\ 
{v}^{\prime }(b,0)&=\dfrac{\lambda (1-c)}{2(\lambda hc-s)}\phi^{0} _{N-1}+\dfrac{\lambda (c-1)}{2(\lambda hc-s)}%
\phi^{0} _{N+1}%
\end{aligned}
\label{B2}
\end{equation}

\section{Simulations of Motions of Patterns Governed by Gray-Scott System}

\noindent
In this section, the diffusion of some initial pulses are obtained by using both the differential quadrature and collocation methods. 

\noindent
In the first example, the replications of two initial pulses formed by initial conditions represented by $u$ and $v$ and covering the related domain are simulated for particular choice of diffusion coefficients. The results of decreasing one of the diffusion coefficients in the system are also examined.

\noindent
In the second example, the initial conditions for both $u(x,t)$ and $v(x,t)$ in the first example are manipulated to give two initial pulses represented by both functions. The diffusion and replicating themselves of more initial pulses, as a result covering the domain is simulated.   

\subsection{Diffusion of Single Initial Pulse} 
\noindent
The diffusion of two single initial pulses represented by $u(x,t)$ and $v(x,t)$ are simulated by the proposed methods. This problem is the solution of the initial-boundary value problem for the system (\ref{gs}) combined with the initial conditions 
\begin{equation}
\begin{aligned}
u(x,0)&=1-\frac{1}{2}\sin ^{100}\left( \pi x\right) \\
v(x,0)&=\frac{1}{4}\sin ^{100}\left( \pi x\right)  
\end{aligned} \label{zegic}
\end{equation}%
and the boundary conditions (\ref{bc})\cite{zeg}. In the first case, the diffusion of initial pulse is studied. The simulations are accomplished in $t \in [0,2000]$ with the parameters $\varepsilon _{1}=10^{-4},$ $\varepsilon _{2}=10^{-6},$ $f=0.024,$ $k=0.06$ over the finite interval $[0,1]$. The time increment $\Delta t=0.1$ is used with various numbers of grids in the interval of the problem. The simulations of diffusion of both $u(x,t)$ and $v(x,t)$ obtained by the proposed methods are plotted in Fig. \ref{fig:1a} and Fig. \ref{fig:1b}. The motions of both pulses can be observed more clearly by tracking the projections, Fig. \ref{fig:1c} and Fig. \ref{fig:1d}. 

\noindent
The initial upside-down pulse represented by the initial condition for $u(x,t)$ replicates itself at the beginning of the simulation, Fig. \ref{fig:1a}. Both replicas begin to travel on opposite sides and get far away from each other, Fig. \ref{fig:1c}. At about $t=900$, both of the pulses replicate themselves again and the number of pulses reaches four. All replicas keep their propagations as time goes. The ones close to the ends of the interval continue to their motions towards the ends of the interval. The remaining two that are close to each other propagate towards each other. 

\noindent
The pulse represented by $v(x,t)$ is a positive pulse initially. At the beginning of the simulation, the initial pulse replicates itself, Fig. \ref{fig:1b}. These two replicas separate from each other and move far away as time goes. At about the time $t=900$ both pulses replicate themselves, Fig. \ref{fig:1d}. The separation of new replicates lasts till approximately $t=1000$. After this time, we observe four well-separated pulses. The inner replicas move closer to each other, as the others moving towards the ends of the domain.  
Both simulations of $u(x,t)$ and $v(x,t)$ generated by both proposed methods are in a good agreement with expected results and the simulations with Zegeling\& Kok\cite{zeg}' s results. 

\begin{figure}[htp]
    \subfigure[Coverage of the domain by the initial single pulse represented by $u$]{
   \includegraphics[scale =0.35] {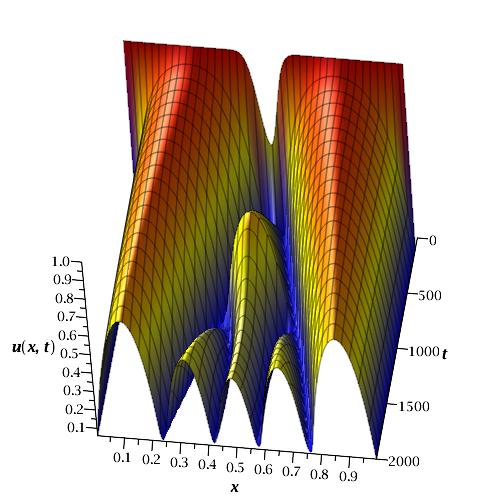}
   \label{fig:1a}
 }
 \subfigure[Coverage of the domain by the initial single pulse represented by $v$]{
   \includegraphics[scale =0.35] {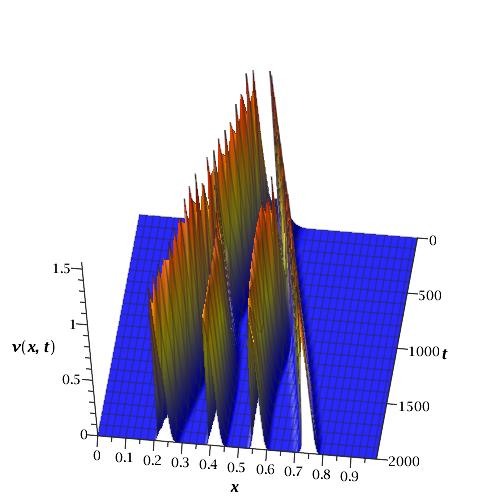}
   \label{fig:1b}
 }
  \subfigure[Projection of $u$]{
   \includegraphics[scale =0.35] {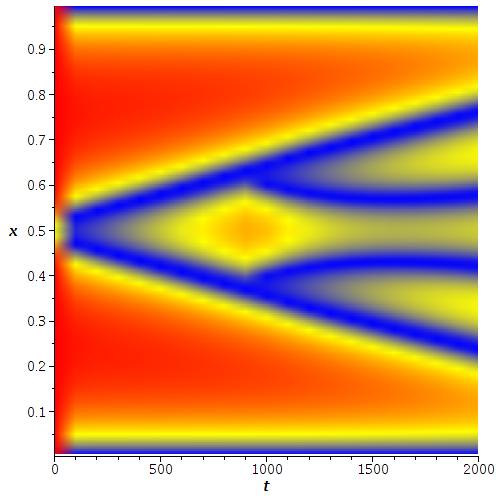}
   \label{fig:1c}
 }
  \subfigure[Projection of $v$]{
   \includegraphics[scale =0.35] {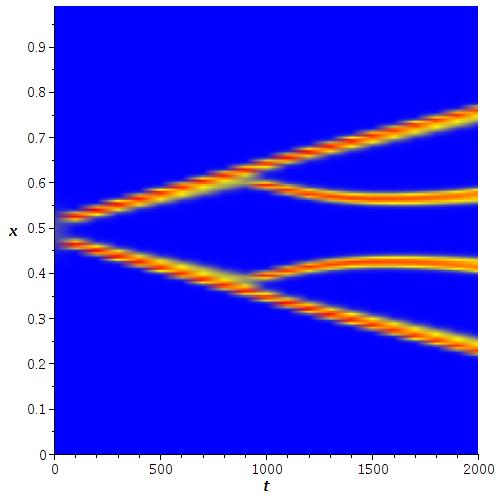}
   \label{fig:1d}
 }
 \caption{Occupation of a domain by single initial pulses in $t\in [0,2000]$}
\end{figure}

\noindent


\noindent
We run the same simulation with the same values of the parameters in the same initial conditions except $\varepsilon _{1}$. 

\noindent
The selection of the diffusion coefficient $\varepsilon _{1}$ as $10^{-5}$ causes that the initial pulses represented by $u(x,t)$and $v(x,t)$ replicates themselves only at the beginning of the simulation. The well-separated pulses represented by $u(x,t)$ protect their positions by standing where they are after replication during the long simulation period. Both pulses preserve their shapes and heights as time goes and do not generate more replicas in the simulation process time. The same results can also be stated for the two replicas of initial pulse represented by the function $v(x,t)$. The replicas are formed at the beginning of the simulation duration. Neither of them move in any direction but both preserve their shapes like the pulses represented by $u(x,t)$. The projections of the simulations of both functions $u(x,t)$ and $v(x,t)$ in the long simulation period $[0,25000]$ are given in Fig \ref{fig:3a}-Fig \ref{fig:3b}.

\begin{figure}[htp]
  \subfigure[Projection of $u$]{
   \includegraphics[scale =0.35] {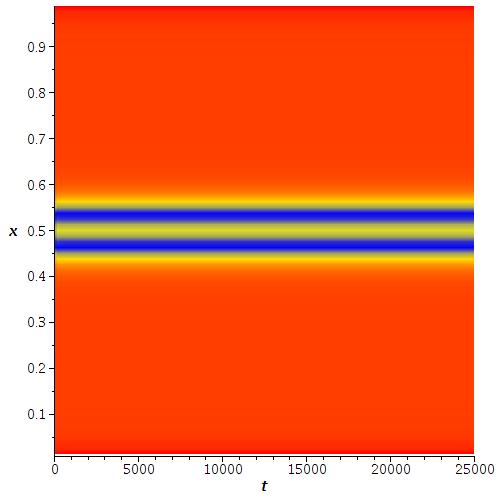}
   \label{fig:3a}
 }
  \subfigure[Projection of $v$]{
   \includegraphics[scale =0.35] {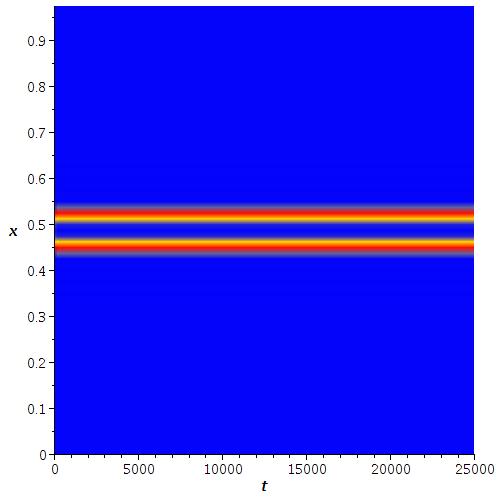}
   \label{fig:3b}
 }
 \caption{Simulations of standing pulses $t\in [0,25000]$}
\end{figure}

\subsection{Diffusion of Double Initial Pulses}

\noindent
The initial conditions (\ref{zegic}) have been manipulated to give two separated initial pulses by rewriting the sine function in the initial conditions as sum of two sine functions. Then, $x-x_0$ and $x-\tilde{x_0}$ are substituted instead of $x$ in the modified initial condition containing the sum of two sine functions. Thus, the peaks of both pulses are positioned at $x_0$ and $\tilde{x_0}$. Under this modifications the initial conditions are formed as
\begin{equation}
\begin{aligned}
u(x,0)&=1-\left[ 0.5 \sin^{100}(\pi (x-x_0))+0.5 \sin^{100}(\pi (x-\tilde{x_0}))\right] \\
v(x,0)&=\frac{1}{4}\sin ^{100}\left( \pi (x-x_0)\right)+\frac{1}{4}\sin ^{100}\left( \pi (x-\tilde{x_0}) \right) \\
\end{aligned}
\end{equation}
for both $u(x,t)$ and $v(x,t)$. In order to prevent the moving pulses as time goes to hit the ends of the problem interval, we decreased the diffusion coefficient $\varepsilon _{1}$ to $5\times 10^{-5}$. The remaining parameters are chosen as the same in the previous example. The numerical solutions are determined in $t \in [0,1200]$ with the time increment $\Delta t=0.1$ over the finite interval $[0,1]$. The numerical solutions simulating diffusion of two initial pulses are plotted in Fig. \ref{fig:2a} and Fig. \ref{fig:2b}. The occupation of the domain as a result of the diffusion can be observed clearly in the projections of both functions Fig. \ref{fig:2c} and Fig. \ref{fig:2d}.

\noindent
For the sake of compatibility, the peaks of two initial upside-down pulses represented by $u(x,0)$ are positioned at $x=0.25$ and $x=0.75$ by choosing $x_0=0.25$ and $\tilde{x_0}=0.75$, Fig. \ref{fig:2a}. Both initial pulses replicate themselves at the beginning of the simulation and separate clearly from each other in the first $100$ unit time, Fig. \ref{fig:2c}. As time goes, four pulses keep their diffusion till $t=700$. When the time reaches $t=700$, the replication process starts again. All four pulses replicate themselves to give eight pulses. All pulses keep to cover the interval until the end of simulation terminating time.  

\noindent
Two positive initial pulses represented by $v(x,t)$ also behave like $u(x,t)$ during the simulation, Fig. \ref{fig:2b}. At the beginning of the simulation, both pulses replicate themselves and diffuse to cover the domain, Fig. \ref{fig:2d}. At about $t=700$, all four pulses replicate themselves again. Totally eight replicas keeps to cover the domain. 
\begin{figure}[htp]
    \subfigure[Diffusion of two initial pulses for $u$]{
   \includegraphics[scale =0.35] {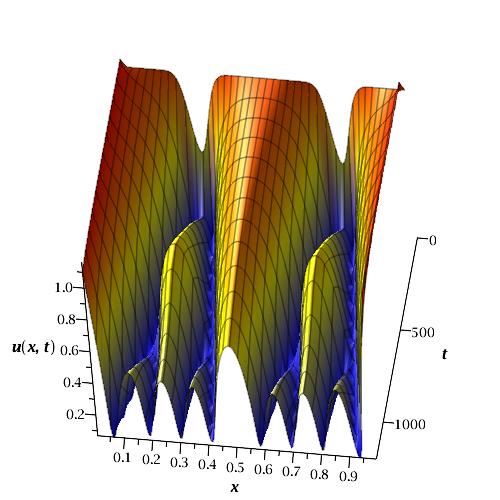}
   \label{fig:2a}
 }
 \subfigure[Diffusion of two initial pulses for $v$]{
   \includegraphics[scale =0.35] {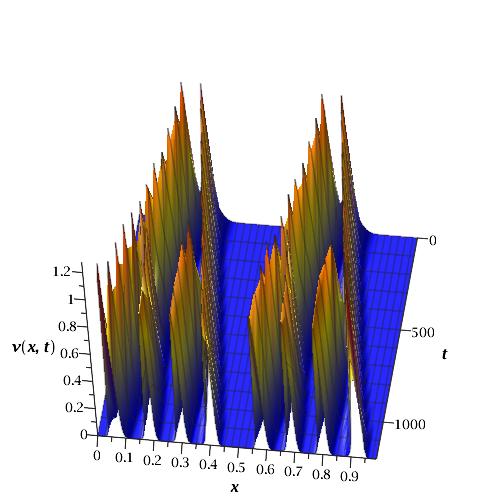}
   \label{fig:2b}
 }
  \subfigure[Projection of $u$]{
   \includegraphics[scale =0.35] {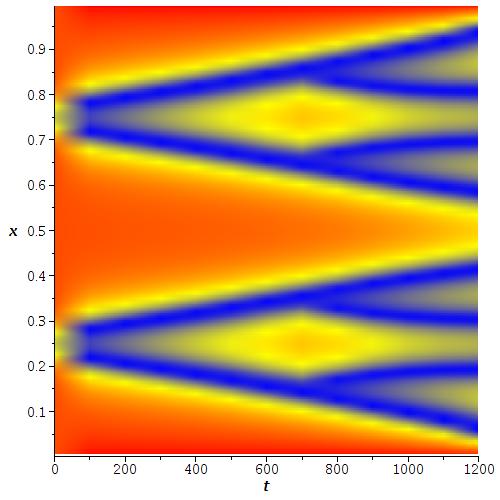}
   \label{fig:2c}
 }
  \subfigure[Projection of $v$]{
   \includegraphics[scale =0.35] {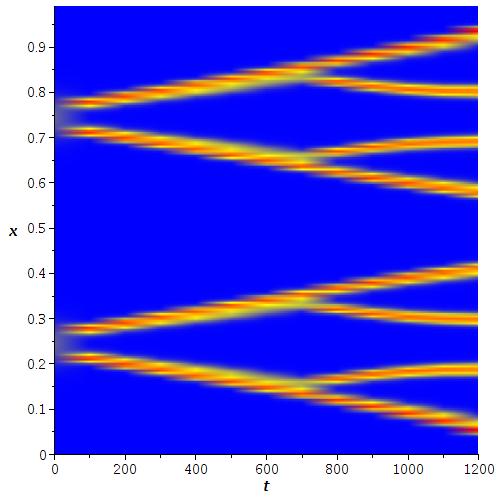}
   \label{fig:2d}
 }
 \caption{Simulations of motion of initial pulses in $t\in [0,1200]$}
\end{figure}
\noindent
\section{Conclusion}

\noindent
In this study, we setup a differential quadrature algorithm combined with implicit Rosenbrock method of order four and a collocation method based on exponential B-spline functions for the numerical solutions of some initial-boundary value problems for a nonlinear Gray-Scott system. 

\noindent
In the first example, we obtained the numerical solutions for the problem modeling single initial pulse occupying the problem interval. The motions and replications of the two initial pulses of the system cover the problem interval as time goes. A particular case for this problem is also simulated by decreasing one of the diffusion coefficient. In this case, the standing replicas do not propagate the problem domain during the simulation process. 

\noindent
In the second example, we perform some manipulations on the initial conditions to construct two well-separated upside-down initial pulses and two positive initial pulses represented by $u(x,0)$ and $v(x,0)$, respectively. The generation of replicas and propagations are simulated successfully. 

\noindent
Both proposed methods have produced acceptable solutions for the Gray-Scott system. Both of the methods remain stable during the long simulation process times.



\begin{thebibliography}{10}
\bibitem{a1} Kondo, S., \& Miura, T. (2010). Reaction-diffusion model as a framework for understanding biological pattern formation. Science, 329(5999), 1616-1620.
\bibitem{tor} Acosta, A.O., Illanes, C. \& Marchese, J. (2009). Removal and recovery of Cr (III) with emulsion liquid membranes.
Desalin. Wat. Treat. 7 (1-3), 18-24.
\bibitem{volki1} Eyupoglu, V., Surucu, A., \& Kunduracioglu, A. (2015). Synergistic extraction of Cr (VI) from Ni (II) and Co (II) by flat sheet supported liquid membranes using TIOA and TBP as carriers. Polish Journal of Chemical Technology, 17(2), 34-42.
\bibitem{volki3} Eyupoglu, V., \& Tutkun, O. (2011). The extraction of Cr (VI) by a flat sheet supported liquid membrane using alamine 336 as a carrier. Arabian Journal for Science and Engineering, 36(4), 529-539.
\bibitem{nicolis} Nicolis, G., \& Prigogine, I. (1977). Self-organization in nonequilibrium systems (Vol. 191977). Wiley, New York.
\bibitem{pear} Pearson, J.E., Complex pattern in a simple system, Science, 261, 189-192, 1993.
\bibitem{lee} Lee, K. J., McCormick, W. D., Pearson, J. E., \& Swinney, H. L. (1994). Experimental observation of self-replicating spots in a reaction-diffusion system. Nature, 369(6477), 215-218.
\bibitem{lee2} Lee, K. J., \& Swinney, H. L. (1995). Lamellar structures and self-replicating spots in a reaction-diffusion system. Physical Review E, 51(3), 1899.
\bibitem{petrov} Petrov, V., Scott, S. K., \& Showalter, K. (1994). Excitability, wave reflection, and wave splitting in a cubic autocatalysis reaction-diffusion system. Philosophical Transactions of the Royal Society of London A: Mathematical, Physical and Engineering Sciences, 347(1685), 631-642.
\bibitem{reynolds} Reynolds, W. N., Pearson, J. E., \& Ponce-Dawson, S. (1994). Dynamics of self-replicating patterns in reaction diffusion systems. Physical review letters, 72(17), 2797.
\bibitem{gray1} Gray, P., \& Scott, S. K. (1984). Autocatalytic reactions in the isothermal, continuous stirred tank reactor: Oscillations and instabilities in the system A+2B $\rightarrow$ 3B; B$\rightarrow$ C. Chemical Engineering Science, 39(6), 1087-1097.
\bibitem{hale} Hale, J. K., Peletier, L. A., \& Troy, W. C. (2000). Exact Homoclinic and Heteroclinic Solutions of the Gray-Scott Model for Autocatalysis. SIAM Journal on Applied Mathematics, 61(1), 102-130.
\bibitem{zeg} Zegeling, P. A., \& Kok, H. P. (2004). Adaptive moving mesh computations for reaction-diffusion systems. Journal of Computational and Applied Mathematics, 168(1), 519-528.
\bibitem{selkov} Sel'Kov, E. E. (1968). Self-oscillations in glycolysis. Eur. J. Biochem, 4(1), 79-86.
\bibitem{gray2} Gray, P., \& Scott, S. K. (1985). Sustained oscillations and other exotic patterns of behavior in isothermal reactions. The Journal of Physical Chemistry, 89(1), 22-32.
\bibitem{doelman} Doelman, A., Gardner, R. A., \& Kaper, T. J. (1998). Stability analysis of singular patterns in the 1D Gray-Scott model: a matched asymptotics approach. Physica D: Nonlinear Phenomena, 122(1), 1-36.
\bibitem{owo1} Owolabi, K. M., \& Patidar, K. C. (2014). Numerical solution of singular patterns in one-dimensional Gray-Scott-like models. International Journal of Nonlinear Sciences and Numerical Simulation, 15(7-8), 437-462.
\bibitem{kuz1} Kuznetsov, Y. A., Meijer, H. G., Al-Hdaibat, B., \& Govaerts, W. (2015). Accurate approximation of homoclinic solutions in Gray-Scott kinetic model. International Journal of Bifurcation and Chaos, 25(09), 1550125.
\bibitem{coo1} Cooper, F., Ghoshal, G., \& Perez-Mercader, J. (2013). Composite bound states and broken U(1) symmetry in the chemical-master-equation derivation of the Gray-Scott model. Physical Review E, 88(4), 042926.
\bibitem{akmaz2} Akmaz, H. K. (2009). Three-dimensional elastic problems of three-dimensional quasicrystals. Applied Mathematics and Computation, 207(2), 327-332.
\bibitem{sahin1} Sahin, A., \& Ozmen, O. (2014). Usage of Higher Order B-splines in Numerical Solution of Fisher's Equation. International Journal of Nonlinear Science, 17(3), 241-253.
\bibitem{aksoy1} Aksoy, A. M., Irk, D., \& Dag, I. (2012). Taylor collocation method for the numerical solution of the nonlinear Schrödinger equation using quintic B-spline basis. Physics of Wave Phenomena, 20(1), 67-79.
\bibitem{sahin2} Sahin, A., Dag, I., \& Saka, B. (2008). A B-spline algorithm for the numerical solution of Fisher's equation. Kybernetes, 37(2), 326-342.
\bibitem{sahin3} Dag, I., Şahin, A., \& Korkmaz, A. (2010). Numerical investigation of the solution of Fisher's equation via the B?spline Galerkin method. Numerical Methods for Partial Differential Equations, 26(6), 1483-1503.
\bibitem{akmaz3} Akmaz, H. K. (2009). Variational iteration method for elastodynamic Green's functions. Nonlinear Analysis: Theory, Methods \& Applications, 71(12), e218-e223.
\bibitem{geyikli1} Geyikli, T., \& Karakoç, S. B. G. (2012). Petrov-galerkin method with cubic B-splines for solving the MEW equation. Bulletin of the Belgian Mathematical Society-Simon Stevin, 19(2), 215-227.
\bibitem{akmaz4} Akmaz, H. K., \& Korkmaz, A. (2012). An analytic solution of initial boundary value problem for 3D quasicrystals in half space. Philosophical Magazine Letters, 92(10), 572-579.
\bibitem{esen1} Esen, A., Ucar, Y., Yagmurlu, N., \& Tasbozan, O. (2013). A Galerkin finite element method to solve fractional diffusion and fractional diffusion-wave equations. Mathematical Modelling and Analysis, 18(2), 260-273.
\bibitem{dursun1} Irk, D., Dag, I., \& Tombul, M. (2015). Extended cubic B-spline solution of the advection-diffusion equation. KSCE Journal of Civil Engineering, 19(4), 929-934.
\bibitem{mccartin} McCartin, B. J. (1991). Theory of exponential splines. Journal of Approximation Theory, 66(1), 1-23.
\bibitem{sakai} Sakai, M., \& Usmani, R. A. (1989). A class of simple exponential B-splines and their application to numerical solution to singular perturbation problems. Numerische Mathematik, 55(5), 493-500.
\bibitem{radu} Radunovic, D. (2008). Multiresolution exponential B-splines and singularly perturbed boundary problem. Numerical Algorithms, 47(2), 191-210.
\bibitem{rao} Rao, S. C. S., \& Kumar, M. (2008). Exponential B-spline collocation method for self-adjoint singularly perturbed boundary value problems. Applied Numerical Mathematics, 58(10), 1572-1581.
\bibitem{reza} Mohammadi, R. (2013). Exponential B-Spline Solution of Convection-Diffusion Equations. Applied Mathematics, 4(06), 933.
\bibitem{ozlem} Ersoy, O., \& Dag, I. (2015). The Exponential Cubic B-Spline Algorithm for Korteweg-de Vries Equation. Advances in Numerical Analysis, 2015.
\bibitem{ozlem1} Ersoy, O., \& Dag, I. (2015). Numerical solutions of the reaction diffusion system by using exponential cubic B-spline collocation algorithms. Open Physics, 13(1).
\bibitem{korkmaz2} A.~Korkmaz, I.~Dag, Shock wave simulations using sinc differential quadrature method, Engineering Computations 28 (2011) 654-674.
\bibitem{bellomo1} Bellomo, N., \& Ridolfi, L. (1995). Solution of nonlinear initial-boundary value problems by sinc collocation-interpolation methods. Computers \& Mathematics with Applications, 29(4), 15-28.
\bibitem{stenger}
F.~Stenger, Numerical Methods Based on Sinc and Analytic Functions, Springer,
  New York, 1993.

\bibitem{carlson1}
T.~S. Carlson, J.~Dockery, J.~Lund, A sinc-collocation method for initial
  boundary value problems, Mathematics of Computation 66 (1997) 215--235.

\bibitem{secer1}
A.~Secer, Numerical solution and simulation of second-order parabolic pdes with
  sinc-galerkin method using maple, Abstract and Applied Mathematics Article ID
  686483 (2013) 1--10.

\bibitem{dehghan2}
M.~Dehghan, A.~Saadatmandi, The numerical solution of a nonlinear system of
  second-order boundary value problems using the sinc-collocation method,
  Mathematical and Computer Modelling 46 (2007) 1434--1441.
\bibitem{lund1}
J.~Lund, K.~L. Bowers, Sinc Methods for Quadrature and Differential Equations,
  SIAM, Philadelphia, 1992.
	\bibitem{bellman1}
R.~Bellman, B.~G. Kashef, J.~Casti, Differential quadrature: A tecnique for the
  rapid solution of nonlinear differential equations, Journal of Computational
  Physics 10 (1972) 40--52.



\end{thebibliography}
\end{document}